# Beneficial Investigation of Extended-Range Electric Powertrains with Dual-Motor Inputs and Multi-Speed Transmission

Cong Thanh Nguyen, Paul D. Walker, Nong Zhang*

School of Mechanical and Mechatronic Engineering, Faculty of Engineering and Information Technology, University of Technology Sydney, 15 Broadway, 2007 Ultimo, NSW, Australia

* Corresponding author: Nong Zhang

Email address: Nong.Zhang@uts.edu.au

## Abstract

This paper comparatively investigates the performance of extended-range electric powertrains composed by integrating dual-motor inputs, multi-speed transmission, and engine in either series or parallel connection. Two configurations, namely dual-motor series powertrain (DMSP) and dual-motor parallel powertrain (DMPP), feature the same electric drivetrain of two downsized motors and a four-speed transmission, but their range extenders are fundamentally different. While the DMSP consists of a range extender formed by an engine-generator unit, the DMPP connects the engine through a frictional clutch. This study starts with the parameter selection that guarantees the equivalent dynamics ability of all configurations. Mathematic models are second established in detail. A model predictive control-based energy management strategy is third presented. Since the recommended configurations aim for bus application, the driving cycle CBDC and ECE15×5 are chosen for simulation. Performance indexes used for comparison include electric consumption, fuel consumption, and emissions (hydrocarbon, carbon monoxide, nitrogen oxides, and particulate matter). Compared to the conventional single-motor series powertrain, the DMSP improves all the indexes significantly, while the DMPP decreases fuel consumption further but increases most noxious exhaust emissions.

# 1. Introduction

With the additional function of on-board electricity generation, extended-range electric vehicles (EREVs) can overcome the current limitations of pure electric vehicles (EVs), i.e., short driving range, high battery cost, charging station requirements, and interruption for plug-in charging. Powertrain configuration is a decisive factor that determines vehicle dynamic and energy efficiency. The EREV configuration can be separated into electric drivetrain and range extender.[1] Since EREVs can be considered as EVs with the function of on-board electricity production, the electric drivetrain configuration can apply all developments of EVs. Firstly, multi-speed transmissions significantly improve both dynamic performance and energy efficiency of electric powertrains.[2, 3] Simulation results showed that two, three, and four-speed transmissions enhance energy efficiency by 5-12% in comparison with the single-speed ratio.[3] Secondly, two downsized-motor inputs operate independently or cooperatively with higher torque utilization factors than the single motor.[4, 5] Consequently, the motors work more frequently in their peak efficiency regions, so they obtain a better energy economy. Due to the increasing demand for drivetrain development, the combinations of dual-motor inputs and multi-speed transmission have been introduced to EVs. For example, an auxiliary motor is added to the output shaft to support the primary motor, which drives in a three-speed transmission.[6, 7] The auxiliary motor mainly operates during the high power demand and gear shifting, so overall efficiency is still limited. Another design is the composition of two identical motors and four-speed transmission.[8, 9] This design not only improves dynamic performance and energy efficiency but also guarantees the shifting smoothness of the multi-speed transmission. The smoothness results from the inclusion of two independent power flows that can compensate for the torque gap mostly during shifting.

The range extender is generally formed by an engine generator unit (EGU), which connects the engine to the driveline in series.[10] In hybrid electric vehicles (HEVs), the engine can be connected in series or parallel.[11, 12] On the one hand, the series connection decouples the engine from driving cycles, so easier to control, peak engine efficiency, and low noise. The main disadvantage is the considerable energy wasted by the conversion of fuel into electricity. On the other hand, with the mechanical link between the engine and driven wheels, the parallel connection guarantees the energy lost by the conversion process to be relatively low. In general, the series HEVs are more efficient in urban driving cycles while the parallel HEVs achieve better fuel economy in highway driving conditions.[13, 14] Besides, Matlab-based simulations

showed that the series HEV emits less hydrocarbon but more carbon monoxide and nitrogen oxides.[14]

The aforementioned studies have shown the advantage of the dual-motor inputs and multi-speed transmission in EVs. Besides, the engine connections in series and parallel have been investigated for HEV application. However, few papers have undertaken a comprehensive study on the integration of the dual-motor inputs, multi-speed transmission, and engine with either series or parallel connection in EREVs. Therefore, this paper is an expansion research on dynamic ability, energy economy, and emissions of EREV powertrains composed by the above components.

Aside from powertrain configurations, energy management strategies (EMSs) are effective solutions to reduce fuel consumption and emissions. Real-time EMSs can be separated into rule-based EMSs and optimization-based EMSs. Since rule-based EMSs mostly rely on expert knowledge, they are challenging to obtain optimal results.[13, 15] Besides, the recommendation and calibration of control rules require a lot of time and investment. Thus, most recent studies have concentrated on optimization-based EMSs, including Pontryagin's minimum principle,[16] equivalent consumption minimization strategy,[11, 17] artificial neural network,[18] model predictive control (MPC),[19, 20] and so on. With the predicted information of driving conditions from the global positioning system, the MPC-base EMSs can nearly obtain global optimization results.[19] Dynamic programming was generally selected as a solver to optimize cost function in the time horizon.[19, 21]

Based on the above literature review, this paper comparatively investigates the performances of EREVs composed by the dual-motor inputs, four-speed transmission, and engine in either series or parallel connection. Two configurations consist of the same electric drivetrain of dual-motor inputs and four-speed transmission, but their range extenders are different. One range extender connects the engine in series – in the form of an EGU. Another layout links the engine in parallel – a clutch is added between the engine and an input shaft. The parameter design guarantees all configurations to achieve equivalent dynamic requirements. An MPC-based EMS with the forward dynamic programming solver is then designed. The above powertrains are studied in comparison with the conventional single-motor four-speed EREV. Performance indexes used for comparison include electric consumption, fuel consumption, and emissions (hydrocarbon, carbon monoxide, nitrogen oxides, and particulate matter).

The remainder of this paper is divided into the following sections. The first section describes operating modes and parameter selections of the EREVs. In the next section, mathematical models are built, including the models of the powertrain system, vehicle dynamics, engine, and battery. An MPC-based energy management strategy is then developed. Simulation results are analysed in the next section, and our conclusions are finally drawn.

## 2. Powertrain configuration and parameter selection

### *2.1. Powertrain configuration*

Integrating dual-motor inputs, multi-speed transmission, and engine in either series or parallel connection, two EREV configurations are presented in Fig. 1 and Fig. 2. Two configurations, namely dual-motor series powertrain (DMSP) and dual-motor parallel powertrain (DMPP), feature the same electric drivetrain which is composed by two downsized motors and a four-speed transmission.[8, 9] The motors respectively connect to odd gears (1$^{st}$ and 3$^{rd}$ gears) and even gears (2$^{nd}$ and 4$^{th}$ gears) through two input shafts. However, the range extenders of the two layouts are different. While the DMSP consists of a range extender formed by an engine-generator unit (Fig. 1), the DMPP connects the engine through a frictional clutch (Fig. 2).[22]

The DMSP can operate in total 6 modes, specifically 3 electric modes and 3 series hybrid modes. The electric modes can involve the motor 1 only, motor 2 only, or two motors cooperatively. When the EGU turns on, the electric modes becomes the series hybrid modes. On the other hand, the DMPP has a total of 8 modes, including 3 electric modes, engine mode, series hybrid mode, and 3 parallel hybrid modes (Tab. 1). The electric modes are similar to the DMSP, while the parallel hybrid modes are the combinations of the engine and motor 1, or motor 2, or both motors.

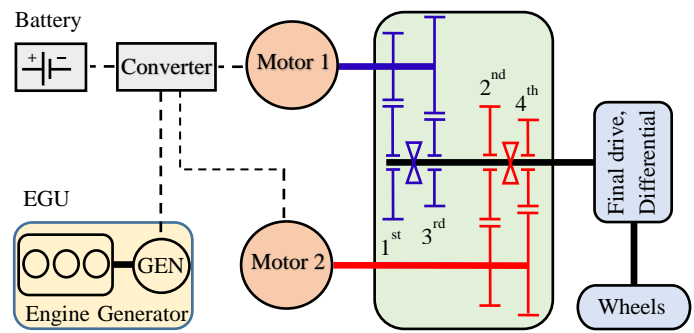

**Fig. 1** Configuration of dual motor-series powertrain (DMSP)

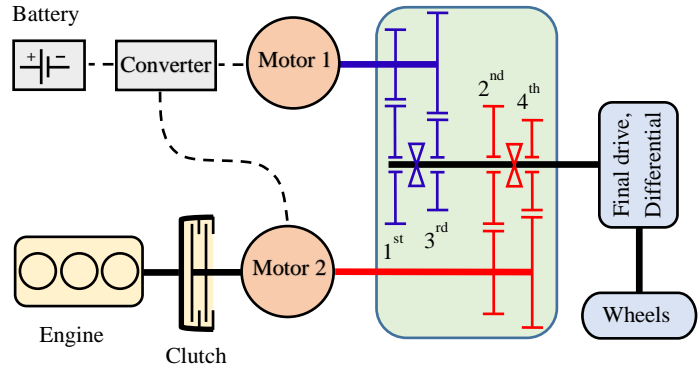

**Fig. 2** Configuration of dual motor-parallel powertrain (DMPP)

**Tab. 1** Operating modes of the DMPP

| Mode | Description | | |
|---|---|---|---|
| | Engine /Clutch | Motor 2/Even Synchronizer | Motor 1/Odd Synchronizer |
| Electric | ×/× | ×/× | ±/● |
| | ×/× | ±/● | ×/× |
| | ×/× | ±/● | ±/● |
| Engine | +/● | ×/● | ×/× |
| Series hybrid | +/● | −/× | ±/● |
| Parallel hybrid | +/● | ×/● | ±/● |
| | +/● | ±/● | ×/× |
| | +/● | ±/● | ±/● |

Description: '±' means positive or negative power; '●' means engaged clutch or synchronizer; '×' means power off, disengaged clutch or synchronizer.

This paper studies the performance of the above configurations in comparison with the conventional single-motor series powertrain (SMSP) (Fig. 3). The SMSP employs an EGU as a range extender. However, its electric drivetrain consists of a single motor (motor 0) and a four-speed transmission.

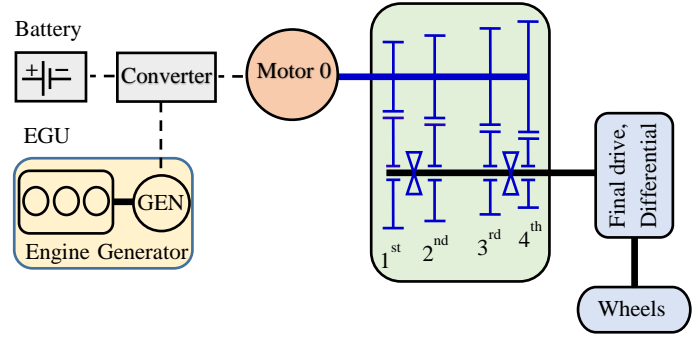

**Fig. 3** Configuration of single motor-series powertrain (SMSP)

*2.2. Parameter selection*

Since parameter optimization is out of the research scope, powertrain parameters are estimated based on the vehicle operation in boundary conditions, namely maximum acceleration, speed, and grade ability. Besides, the parameters of the SMSP will be calculated in detail to illustrate the general principle of the design process. The parameters of the DMSP and DMPP will be then determined to guarantee the equivalent dynamic performances of all configurations.

Acceleration and grade ability are the base for determining motor power. For acceleration performance, the power to speed up from zero to $v_f$ in the duration of $t_a$ is obtained as:[23]

$$P_{m0}^{acc} = \frac{\delta M}{2t_a}(v_f^2 + v_b^2) + \frac{2}{3}Mgf_r v_f + \frac{1}{5}\rho_a C_d A_f v_f^3 \tag{1}$$

Where $M$ is the gross mass of vehicle, $\delta$ is the combined rotational inertia coefficient, $t_a$ is the acceleration time, $v_f$ is the final speed, $v_b = v_f/i_m$ is the base speed, $i_m$ is the speed ratio of motor, $g$ is the gravitational acceleration, $f_r$ is the tyre rolling resistance, $C_d$ is the aerodynamic drag coefficient, $\rho_a$ is the air density, $A_f$ is the vehicle front area.

For grade ability, the power to propel the vehicle in the maximum inclination can be expressed as:

$$P_{m0}^{grade} = \left(Mg \sin \varphi_{max} + Mgf_r \cos \varphi_{max} + \frac{1}{2}\rho_a C_d A_f v_\varphi^2\right)v_\varphi \tag{2}$$

The peak power of motor 0 is selected as:

$$P_{m0} = max(P_{m0}^{acc}, P_{m0}^{grade}) \tag{3}$$

Since the motors of the DMSP and DMPP connect to the interlacing gears, the motors are set to be the same size and have a haft power of motor 0. The detail specifications and efficient maps of the motors are presented in Tab. 3 and Fig. 4a, b, respectively.

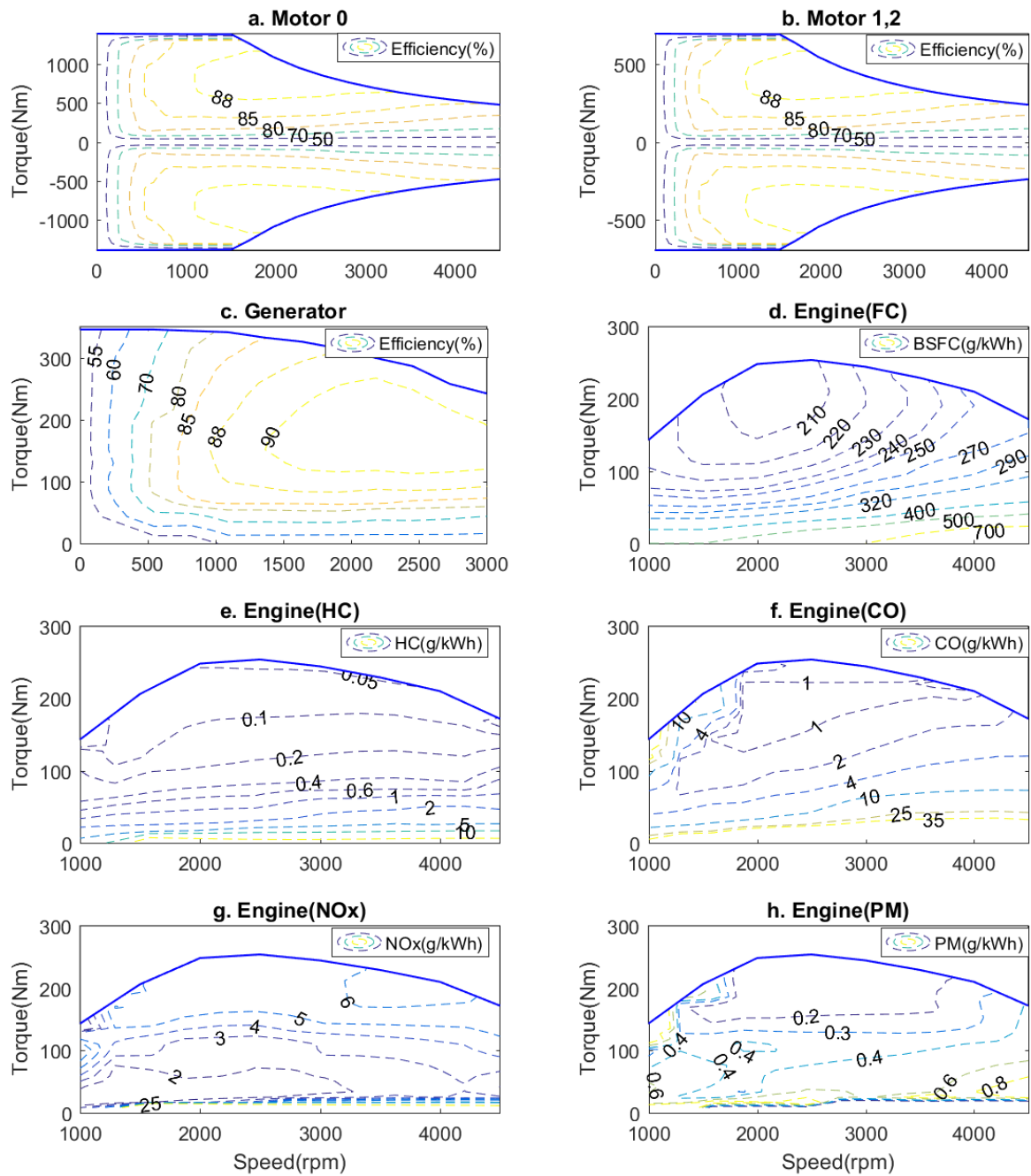

**Fig. 4** Operating maps of the motor, generator, and engine[24]

The engine selection in EREVs depends on battery size, driving range per charge, and control strategy. With the primary purpose of configuration comparison, a 2.5L diesel engine is chosen for all configurations. The generator of the SMSP and DMSP is selected to ensure both the engine and generator to work in their peak efficiency regions. The detail specifications of the engine and generator are presented in Tab. 3. The efficient map of the generator, fuel consumption and emission maps of the engine are shown in Fig. 4c-h.[24]

The four-speed transmission and final drive are designed to guarantee vehicle operation at maximum grade and speed. The vehicle acceleration is presumed to be zero in these conditions. For this reason, the vehicle load torque includes of rolling resistance, inclination resistance, and aerodynamic drag as:

$$T_{load} = \left(Mg \sin \varphi + Mgf_r \cos \varphi + \frac{1}{2}\rho_a C_d A_f v^2\right) r_t \tag{4}$$

Where is $\varphi$ is inclination angle, $r_t$ is the wheel radius.

The lowest gear ratio is identified at the grade of 50% and the constant speed of 10 km/h as:

$$T_{m0}^{max}(v_{\varphi max}) i_1 i_0 \eta_{mw} \geq T_{load}(\varphi_{max}, v_{\varphi max}) \tag{5}$$

Where $\varphi_{max}$ is the maximum inclination angle, $v_{\varphi max}$ is the corresponding speed at the angle of $\varphi_{max}$, $i_1$ is the 1st gear ratio, $i_0$ is the gear ratio of final drive, $T_{m0}^{max}(v_{\phi max})$ is the corresponding torque of motor 0 at the speed $v_{\varphi max}$, $\eta_{mw}$ is the transmission efficiency from motor 0 to the driven wheels.

The highest speed of 80 km/h corresponds to the maximum motor speed and the highest gear ratio (4th gear) as:

$$i_4 i_0 \leq \frac{3.6\pi r_t N_{max}}{30 v_{max}} \tag{6}$$

Where $N_{max}$ is the maximum speed of motor 0.

The traction torque at the highest speed needs to be rechecked as:

$$T_{m0}^{max}(v_{max}) i_4 i_0 \eta_{mw} \geq T_{load}(\varphi_{vmax}, v_{max}) \tag{7}$$

Using the constraints (5-7), the ratio of the final drive, maximum and minimum ratios of the multi-speed transmission can be determined. Besides, the geometrical gear step is used to select the ratios of the intermediate gears.[25]

The principle to select the gear ratios of the DMSP and DMPP is similar to the SMSP. However, both the motors will operate at the maximum inclination and acceleration while only motor 2 will propel in the 4th gear at the highest velocity. Finally, the gear ratios and dynamic performances of all configurations are shown in Tab. 2. The results consist of the maximum vehicle speed ($v_{max}$), grade ability ($\varphi_{max}$), and velocity after 10s acceleration ($v_{max}^{10s}$). The calculation results show that the SMSP and DMSP perform similarly in all conditions. Moreover, with the engine support, the DMPP can achieve better grade and acceleration capability.

**Tab. 2** Dynamic performances of three configurations

|  | SMSP | DMSP | DMPP |
|---|---|---|---|
| $[i_1/i_2/i_3/i_4]$ | 5.0/3.8/2.8/2.1 | 5.9/4.2/3.0/2.1 | |
| $i_0$ | 5.2 | | |
| $v_{max}(km/h)$ | 81.7 | | |
| $\varphi_{max}(\%)$ | 47.3 | 47.2 | 53.1 |
| $v_{max}^{10s}(km/h)$ | 56.6 | 56.3 | 61.8 |

The battery includes 200 lithium-ion cells in series connection. Each cell has the capacity of 180Ah, the nominal voltage of 3.4V. Battery internal resistances and open circuit voltage are presented in Fig. 5 [24]. The parameters of the battery pack are shown as in Tab. 3.

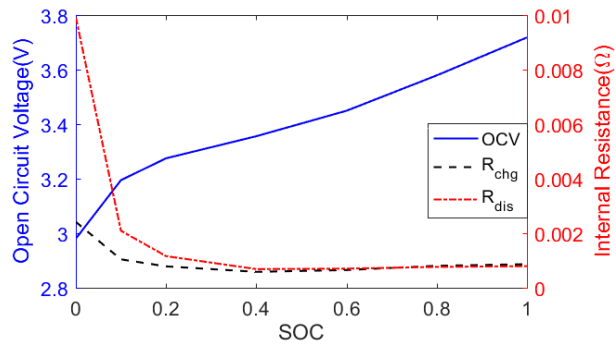

**Fig. 5** Battery internal resistances and open circuit voltage [24]

**Tab. 3.** Specifications of the vehicle and powertrain components

|  | Parameters | Value |
|---|---|---|
| **Vehicle** | Gross weight | 15000 kg |
| **Battery** | Battery type | Lithium ion |
|  | Nominal voltage/Capacity | 680 V/180 Ah |
| **Engine** | Engine type | Diesel, 2.5L |
|  | Maximum power/Maximum speed | 88 kW/4500 rpm |
| **Motors** | Motor type | PMSM |
|  | Peak powers of motor 0/1/2 | 224/112/112 kW |
|  | Max speed/Base speed | 4500/1500 rpm |
| **Generator** | Generator type | PMSM |
|  | Peak power | 76 kW |
|  | Max speed/Base speed | 3000/1000 rpm |

## 3. Mathematical model

### 3.1. Powertrain model

The motors and/or engine supply the required torque to the driven wheels during the acceleration process. In contrast, the motors regenerate braking energy to save in the battery. The mechanical braking system supports the motors to decelerate the vehicle safely. The required torque in the driven wheels of the SMSP, DMSP, and DMPP are sequentially presented as in (8)-(10):

$$T_w = T_{m0} i_G i_0 \eta_t + T_{bF} \tag{8}$$

$$T_w = (T_{m1} i_{Odd} + T_{m2} i_{Even}) i_0 \eta_t + T_{bF} \tag{9}$$

$$T_w = (T_{m1} i_{Odd} + T_{m2} i_{Even} + T_e i_{Even}) i_0 \eta_t + T_{bF} \tag{10}$$

Where $T_{mi}, T_e$ are the torque of the motor $i$ ($i = 0,1,2$) and engine; $i_G, i_{Odd}, i_{Even}$ are the gear ratio of the multi-speed transmission, odd gear, and even gear, respectively; $\eta_t$ is the efficiency of the transmission system; $T_{bF}$ is the front braking torque which equals zero in driving condition.

### 3.2. Vehicle dynamic model

From the longitudinal dynamics, the required torque is consisted of the acceleration torque and load torque as:

$$T_w = \left(\delta M \frac{dv}{dt} + Mg\sin\varphi + Mgf_r\cos\varphi + \frac{1}{2}\rho_a C_d A_f v^2\right) r_t \qquad (11)$$

The total braking torque is generally allocated to all wheels to fully use the tire-to-road adhesion as well as improve vehicle stability. The ideal allocation of the front and rear braking forces (Fig. 6) is expressed as:[26]

$$F_{xR} = \frac{Mg}{2h_g}\left(\sqrt{b^2 + \frac{4h_g L}{mg}F_{xF}} - b\right) - F_{xF} \qquad (12)$$

Where $F_{xF}, F_{xR}$ are the braking forces of the front wheels and rear wheels ($T_{bF} = F_{xF}r_t$, $T_{bR} = F_{xR}r_t$.), $h_g$ is the height of the gravitational centre, $L$ is the wheel base, $b$ is the distance from the gravitational centre to the rear axle.

Since only the rear braking power can be recovered, the ideal allocation wastes a large amount of energy in the front wheels. Therefore, the ideal torque distribution is only applied in the emergency braking process. In the slight and medium braking requirement, using only rear braking torque can satisfy braking requirements and maximize the energy recovery. Consequently, the actual braking distribution is detailed in Fig. 6.

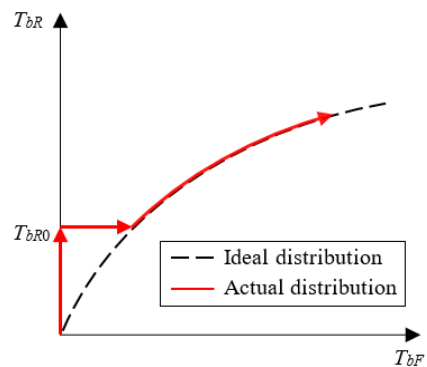

**Fig. 6** Distribution of braking torques[26]

### 3.3. Engine model

Since the engine dynamics is out of this research, the simplified engine model mainly concerns the fuel consumption and emission as:

$$\frac{d(FC)}{dt} = \frac{\mu_e}{\rho_f} P_e \qquad (13)$$

$$\frac{d(GE)}{dt} = \mu_{EG} P_e \quad (GE = \{HC, CO, NOx, PM\}) \qquad (14)$$

Where $FC$ is the fuel consumption, $GE$ is the gas emission (HC, CO, NOx, and PM), $\mu_e$ is the fuel consumption rate, $\rho_f$ is the fuel density, and $\mu_{EG}$ is the emission rate, and $P_e$ is the engine power. The fuel consumption rate and emission rate are interpolated from the engine maps in Fig. 4.

*3.4. Battery model*

This research neglects the effect of battery temperature and age. The battery model is simplified using the open-circuit voltage in series. Therefore, the battery SOC is computed as:[27]

$$\frac{d(SOC)}{dt} = -\frac{\left(U_{OC} - \sqrt{U_{OC}^2 - 4P_{bat}R_{bat}}\right)}{2C_{bat}R_{bat}} \qquad (15)$$

Where $SOC$ is the battery state of charge, $U_{OC}$ is the open circuit voltage, $R_{bat}$ is the internal resistance, $C_{bat}$ is the battery capacity, $P_{bat}$ is the battery power, that is calculated for the SMSP as in (16) and for the DMSP and DMPP as in (17):

$$P_{bat} = P_{m0}(\eta_{m0}\eta_c)^{-sign(P_{m0})} \qquad (16)$$

$$P_{bat} = \left(P_{m1}\eta_{m1}^{-sign(P_{m1})} + P_{m2}\eta_{m2}^{-sign(P_{m2})}\right)\eta_c^{-sign\left(P_{m1}\eta_{m1}^{-sign(P_{m1})} + P_{m2}\eta_{m2}^{-sign(P_{m2})}\right)} \qquad (17)$$

Where $P_{mi}$ and $\eta_{mi}$ are the power and efficiency of the motor $i$ ($i = 0,1,2$), $\eta_c$ is the converter efficiency.

## 4. Energy management strategy

The objective of the energy management strategy is to minimize fuel consumption, emissions and guarantee other dynamic performances. In this research, the EMS considers the control system at the supervisory level that calculates the optimum values of powertrain operation. The optimum values are then used as references for the tracking control in the low-level control. The EMS at the supervisory level is a nonlinear optimization problem that is solved here by the model predictive control (Fig. 7).[28] In each time step, the optimal control sequences are determined to minimize the cost function in the prediction time horizon. The first control sequence is then chosen to apply to the EREV model. The inputs of the MPC controller consists of the required vehicle speed, torque, and other states. The demand speed and torque can be obtained from the driving cycle and calculation, respectively. The other states are the outputs of the EREV model, such as the SOC, gear state,… In practical application, the inputs can be provided by the driver model or by using sensors and observers. The outputs are different for each powertrain configuration and are detailed as follows.

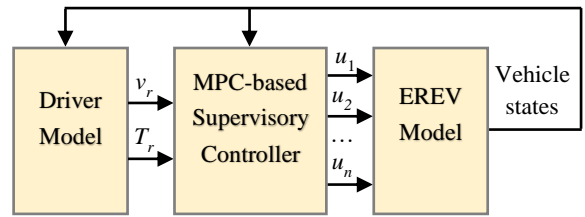

**Fig. 7** Framework of the MPC-based energy management strategy

From the previous equations, the control system is modelled as:

$$\begin{cases} \dot{x} = f(x, u, w) \\ y = g(x, u, w) \end{cases} \quad (18)$$

Where $x = SOC$ is the state variable, $y$ is the measured output, $u$ is the control input, $w = [v_r, T_r]^T$ is the disturbance. The selection of the control input and measured output depends on the powertrain configuration. Therefore, the input and output for the SMSP, DMSP, and DMPP are respectively presented as in (19), (20), and (21):

$$u = [\omega_e, T_e, T_{m0}]^T, y = [FC, GE, \omega_{m0}, T_g, \omega_g, i_{AMT}]^T \quad (19)$$

$$u = [\omega_e, T_e, T_{m1}, T_{m2}]^T, y = [FC, GE, \omega_{m1}, \omega_{m2}, T_g, \omega_g, i_{Odd}, i_{Even}]^T \quad (20)$$

$$u = [T_e, T_{m1}, T_{m2}]^T, y = [FC, GE, \omega_e, \omega_{m1}, \omega_{m2}, i_{Odd}, i_{Even}]^T \quad (21)$$

The EMS requires to use the battery electricity reasonably for the whole driving cycle. However, the optimization in the short time horizon is unable to get global solutions. Therefore, assuming that the required electricity is related to the vehicle power demand, a SOC reference is defined according to the intended driving cycle.[29, 30] The SOC reference will be then added to the cost function for tracking control.

Since electricity is much cheaper than fuel, the battery is used as much as possible at the end. Therefore, the EMS is trivial if the battery can cover all the expected energy. Otherwise, the EMS needs to balance between electricity and fuel consumption. The proposed objective function here mainly consists of fuel consumption and SOC penalty. Besides, the torque fluctuations and state changes of the engine and motors are penalized to guarantee the practical operation. Consequently, the cost functions of the MPC are expressed as in (23) for the SMSP and in (24) for the DMSP and DMPP.

$$J = \sum_{k=m}^{m+p} (FC(k) + \rho_{SOC}|SOC(k) - SOC_r(k)| + \rho_{Se}|S_e(k) - S_e(k-1)| \\ + \rho_{Te}|T_e(k) - T_e(k-1)| + \rho_{Sm0}|S_{m0}(k) - S_{m0}(k-1)| \\ + \rho_{Tm0}|T_{m0}(k) - T_{m0}(k-1)|) \tag{23}$$

$$J = \sum_{k=m}^{m+p} (FC(k) + \rho_{SOC}|SOC(k) - SOC_r(k)| + \rho_{Se}|S_e(k) - S_e(k-1)| \\ + \rho_{Te}|T_e(k) - T_e(k-1)| + \rho_{Sm1}|S_{m1}(k) - S_{m1}(k-1)| \\ + \rho_{Tm1}|T_{m1}(k) - T_{m1}(k-1)| + \rho_{Sm2}|S_{m2}(k) - S_{m2}(k-1)| \\ + \rho_{Tm2}|T_{m2}(k) - T_{m2}(k-1)|) \tag{24}$$

Where $p$ is the prediction time horizon, $S_i (i = e, m0, m1, m2)$ is the state of the engine, motor 0, motor 1, or motor 2 ($S_i = 1$ if the component is on, $S_i = 0$ if the component is off), $\rho_i (i = SOC, Se, Sm0, Sm1, Sm2, Te, Tm0, Tm1, Tm2)$ is the penalty factor.

Then the optimal problem is constructed as:

$$\min_{\substack{SOC_{min} \leq SOC \leq SOC_{max} \\ \omega_i^{min} \leq \omega_i \leq \omega_i^{max}, T_i^{min} \leq T_i \leq T_i^{max}}} J(u) \tag{25}$$

This research uses the forward dynamic programming as an MPC solver which is illustrated in Fig. 8. The battery SOC is selected as a state variable that changes in a narrow range in the time horizon of 5s. As a result, the possible SOC range in each step can be divided into a small grid, while the number of the SOC element is limited. Therefore, the DP solver can calculate the optimal results accurately in the acceptable time requirement. At time point $m$, the MPC uses the forward DP solver to determine the control sequences $[u_1 \; u_2 \ldots u_5]^T$ in the time horizon (Fig. 8). Then, the first control sequence $u_1$ will be selected to apply to the EREV model.

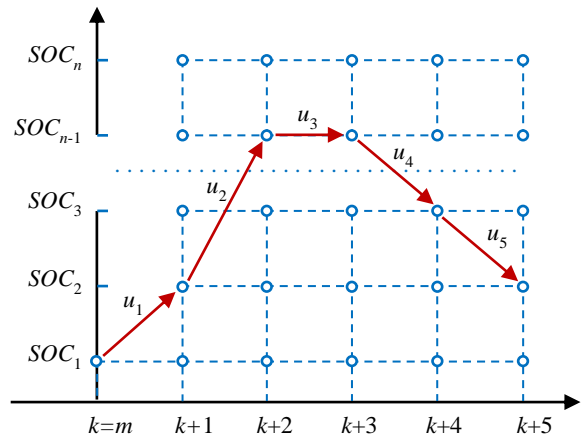

**Fig. 8** Architecture of the dynamic programming solver

## 5. Simulation results

Simulation models are developed in Matlab and are based on the mathematic model and energy management strategy in the previous sections. Since the investigated configurations aim for bus application, the powertrain performance is assessed in two driving cycles, specifically the Chinese bus driving cycle (CBDC) (Fig. 9)[31] and the urban driving cycle (ECE15) (Fig. 13).[32] Besides, depending on the battery capacity and travel length, the operation of the EREVs can be divided into two conditions: electric driving and hybrid driving. In electric driving, motors drive the vehicle without engine support, which results in the identical operation of the DMSP and DMPP. In hybrid driving, the engine keeps the role in generating electricity in all configurations or mechanically drives the vehicle in the DMPP.

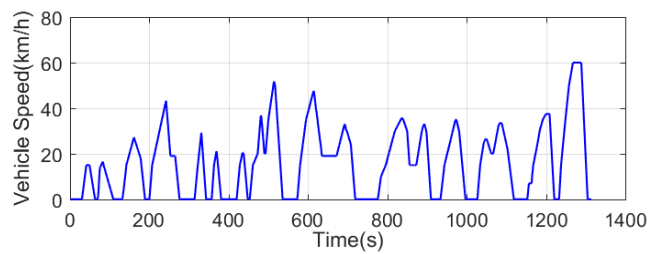

**Fig. 9** Chinese bus driving cycle

Fig. 10 compares the powertrain performance in the electric driving condition in the CBDC. Firstly, gear states are selected from optimal maps,[22] with the gearshifts need to be limited (Fig. 10a, b). Secondly, the motor torques are required to avoid oscillations with large magnitudes

(Fig. 10c, d). In the efficiency map, the working point is determined by the motor speed and torque. With two motor sizes, the left y-axis corresponds to the toques of motor 1 and 2, while the right y-axis corresponds to the torque of motor 0 (Fig. 10e). It is worth noting that motor 1 and motor 2 work in their peak efficiency regions much more frequent than motor 0. Hence, the efficiencies of the dual motors are improved significantly. Especially, motor 0 operates at a mean efficiency of 72.8%, while motor 1 and motor 2 acquire the means of 81.8% and 76.3%, respectively. Finally, Fig. 10f emphasizes the significant reduction of 7.4% in electricity consumption of the DMSP and DMPP (7.18 kWh) in comparison with the conventional SMSP (7.75kWh).

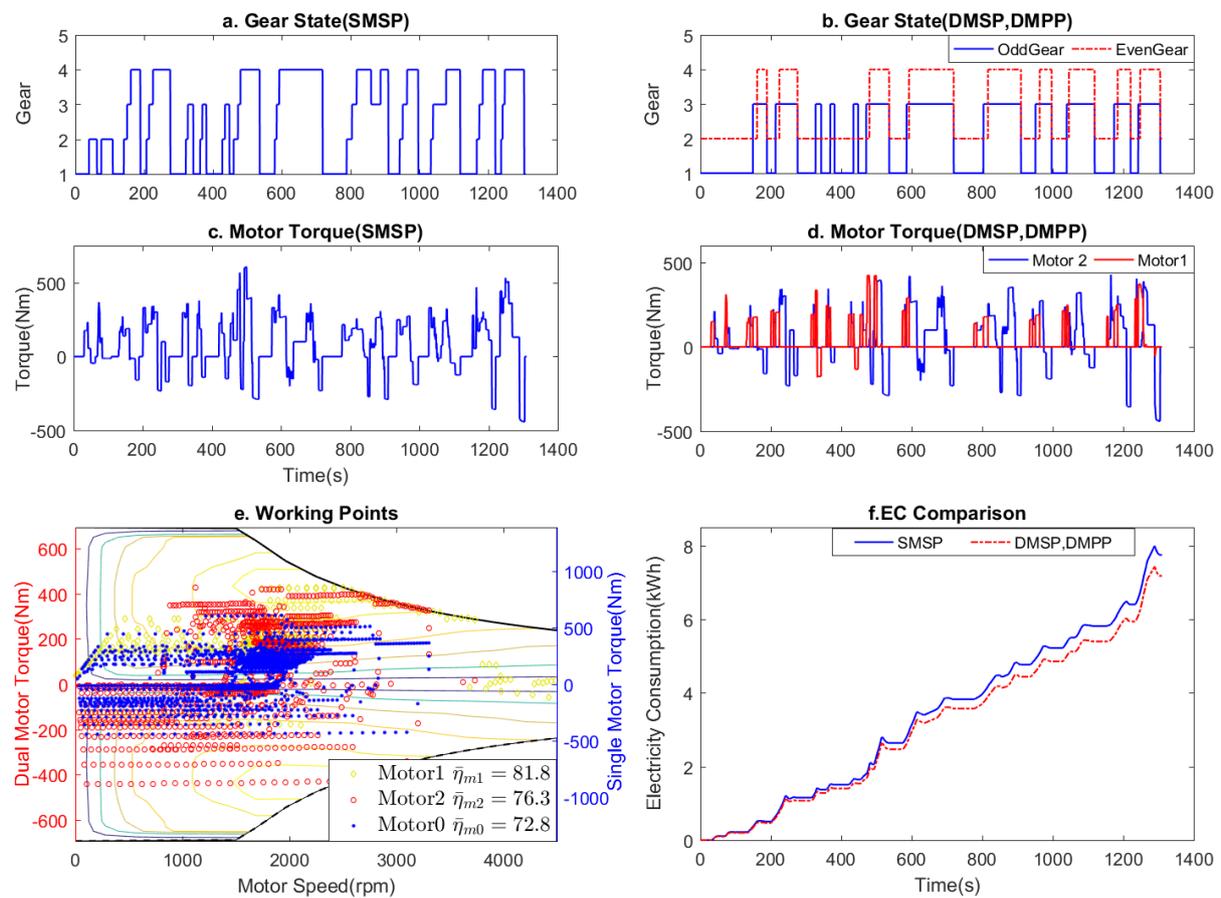

**Fig. 10** Operation comparison in electric driving condition

Fig. 11 compares the EREV performance in the hybrid driving condition in the CBDC. The consumption of fuel and electricity here depends on the rest of battery SOC and travel length. This figure investigates powertrain operation when fuel is the primary energy. Therefore, the battery SOC is set at a low range from 0.32 to 0.3. The proposed SOC reference is shown in Fig. 11e. With the engines connected in series, the motors and transmissions of the SMSP and DMSP operate similarly as in the electric driving condition. Besides, the engines of the SMSP and DMSP work with steady torques (Fig. 11a, b) around their peak efficiency regions (Fig. 11g). For the DMPP, the engine can propel the vehicle mechanically or drive the dual motors to generate electricity (Fig. 11c, d). Therefore, the engine is required to operate with high torque variations (Fig. 11c) and its working points are in a wider region (Fig. 11h). The SOC tracking control shows all final values close to the minimum of 0.3 (Fig. 11e). Consequently, the energy economy is determined by the fuel consumption (Fig. 11f). It is evident that the fuel consumption reduces as in the following sequence: the SMSP, DMSP, and DMPP. The statistical results are synthesized later.

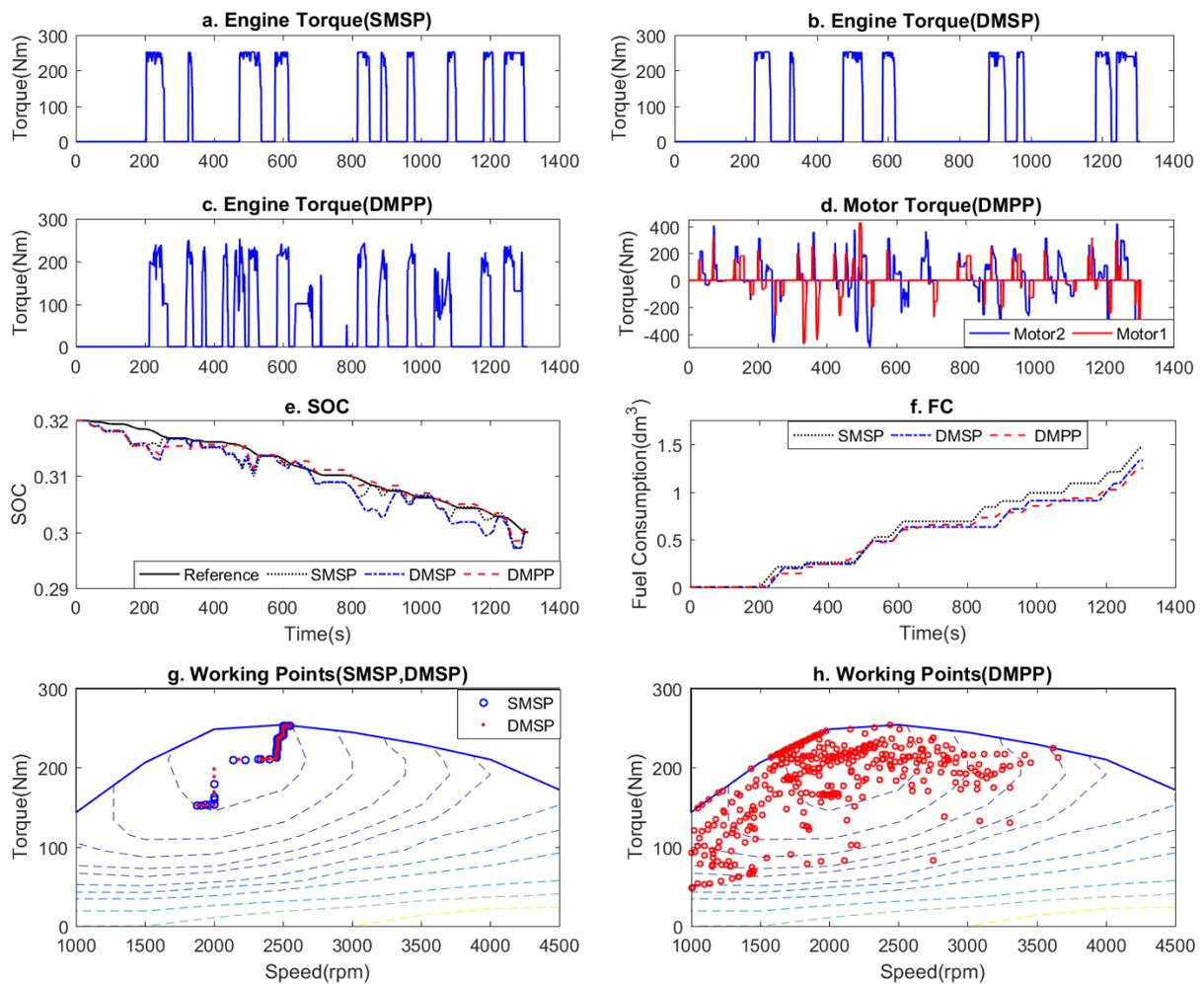

**Fig. 11** Operation comparison in hybrid driving condition

Fig. 12 compares the engine emissions in the hybrid driving condition in the CBDC. The result shows clearly that the DMPP emits more HC, CO, and PM, but less NOx than the SMSP and DMSP. In addition, the emissions of the DMSP are always less than of the conventional SMSP.

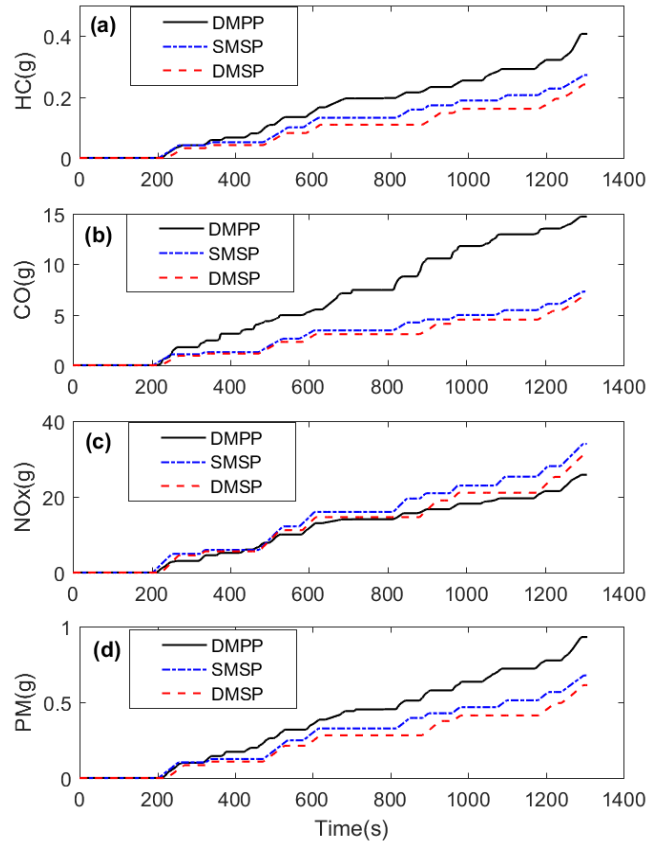

**Fig. 12** Comparison of exhaust emissions in hybrid driving condition

Tab. 4 statistically details the performance indexes used to justify the effectiveness of the recommended EREVs in the CBDC. The performance indexes consist of electric consumption (EC), fuel consumption (FC), and gas emissions (hydrocarbon (HC), carbon monoxide (CO), nitrogen oxides (NOx), and particulate matter (PM)). The indexes of the DMSP and DMPP are compared to the conventional SMSP. In the electric driving condition, the DMSP and DMPP reduce EC significantly by 7.4%. In the hybrid driving condition, the FC reductions are more considerable, by 8.9% for the DMSP, and 14.4% for the DMPP. For emission comparison, the DMSP improves all emission indexes by reducing from 7.4% in CO to 10.3% in PM. On the other hand, the DMPP only reduces NOx by 24.0% but increases the other gases remarkably by 85.7% HC, 102.1% CO, and 36.8% PM. The high emissions of the DMPP are explained by matching the engine operating points (Fig. 11h) to the emission maps (Fig. 4). It clearly shows that the engine of the DMPP frequently runs in high emission regions than the DMSP.

**Tab. 4** Comparison of performance indexes in the CBDC

| Driving condition | Performance index | Configuration | | | Reduction (%) | |
|---|---|---|---|---|---|---|
| | | SMSP | DMSP | DMPP | DMSP | DMPP |
| Electric driving | EC (kWh) | 7.75 | 7.18 | 7.18 | **+7.4** | **+7.4** |
| Hybrid driving | FC (dm$^3$) | 1.46 | 1.33 | 1.25 | **+8.9** | **+14.4** |
| | HC (g) | 0.21 | 0.19 | 0.39 | **+9.5** | **−85.7** |
| | CO (g) | 7.29 | 6.75 | 14.73 | **+7.4** | **−102.1** |
| | NOx (g) | 33.97 | 30.91 | 25.83 | **+9.0** | **+24.0** |
| | PM (g) | 0.68 | 0.61 | 0.93 | **+10.3** | **−36.8** |

The performance indexes are investigated in another urban driving cycle-ECE15. Since the ECE15 is short, the simulation is conducted in five continuing cycles, namely ECE15×5 (Fig. 13). In general, the calculation results in the ECE15×5 are consistent with the outcomes in the CBDC (Tab. 5). Compared to the SMSP, the DMSP reduces significantly 8.3% EC and 11.5% FC in the electric and hybrid driving, respectively. The emission reductions are more considerable, from 11.6% in NOx to 25% in HC. For the DMPP, the energy economy is even better than the DMSP, with the decreases of 8.3% EC and 14.5% FC. Nevertheless, the emissions increase considerably in HC (20%), CO (47.6%), and PM (9.7%), except for NOx.

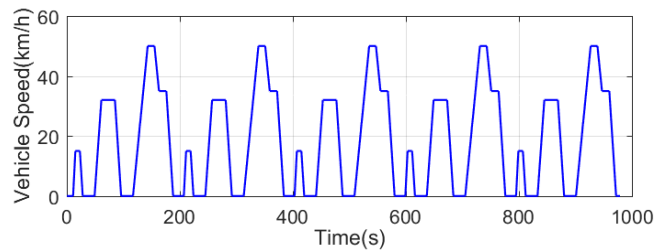

**Fig. 13** ECE15×5 driving cycle

Tab. 5 Comparison of performance indexes in the ECE15×5

| Driving condition | Performance index | Configuration | | | Reduction (%) | |
|---|---|---|---|---|---|---|
| | | SMSP | DMSP | DMPP | DMSP | DMPP |
| Electric driving | EC (kWh) | 7.27 | 6.67 | 6.67 | **+8.3** | **+8.3** |
| Hybrid driving | FC (dm$^3$) | 1.31 | 1.16 | 1.12 | **+11.5** | **+14.5** |
| | HC (g) | 0.20 | 0.15 | 0.24 | **+25.0** | −20.0 |
| | CO (g) | 6.51 | 5.21 | 9.61 | **+20.0** | −47.6 |
| | NOx (g) | 30.29 | 26.79 | 24.25 | **+11.6** | **+19.9** |
| | PM (g) | 0.62 | 0.50 | 0.68 | **+19.4** | −9.7 |

## 6. Conclusion

This paper investigates the performance of two EREV configurations, namely dual-motor series powertrain (DMSP) and dual-motor parallel powertrain (DMPP). The DMSP can operate in total 6 modes in comparison with 8 modes of the DMPP. Parameter selection of all powertrain components are based on dynamic performances, i.e., maximum speed, acceleration, and grade ability. Mathematic models are then presented in detail. A model predictive control-based energy management strategy are next developed. The strategy prioritizes to minimize energy consumption while guarantees the engine and motors to change their torques and states reasonably. The recommended powertrains are finally simulated in comparison with the conventional single-motor series powertrain (SMSP) in the driving cycle CBDC and ECE15×5. Two driving conditions corresponding to without and with engine support are conducted. Performance indexes used for evaluation consist of electric consumption, fuel consumption, and gas emissions.

Simulation results show that two downsized motors operate in their peak efficiency regions more frequently than the single motor. Therefore, compared to the SMSP, both the DMSP and DMPP reduce the electricity and fuel consumption significantly. Depending on the driving cycle and driving condition, the DMSP and DMPP reduces energy usage up to 11.5% and 14.5%, respectively. Furthermore, the DMSP reduces all exhaust emissions considerably, for instance, from 11.6% in NOx to 25% in HC in the ECE15×5. Conversely, the DMPP increases almost emissions except for NOx. In summary, the dual motor powertrains significantly improve the energy economy of the EREVs. Compared to the DMSP, the DMPP decreases fuel consumption further but increases almost exhaust emissions.

# Acknowledgement

This project was supported by the Australian Research Council under Discovery Early Career Researcher Award (DE0170100134), and University of Technology Sydney.